\documentclass[amsmath,prl,
showpacs, elsarticle, twocolumn] {revtex4}
\usepackage{bm}
\usepackage{enumerate}
\usepackage{graphicx}
\usepackage[dvips]{epsfig}
\usepackage{epsf}
\usepackage{xcolor}
\makeatletter
\newcommand{\beq}{\begin{equation}}
\newcommand{\eeq}{\end{equation}}
\newcommand{\bea}{\begin{eqnarray}}
\newcommand{\eea}{\end{eqnarray}}
\newcommand{\bse}{\begin{subequations}}
\newcommand{\ese}{\end{subequations}}
\newcommand{\bwt}{\begin{widetext}}
\newcommand{\ewt}{\end{widetext}}
\newcommand{\Rmnum}[1]{\expandafter\@slowromancap\romannumeral #1@}

\makeatletter

\begin{document}
\title{In-plane Hall effect in two-dimensional helical electron systems
}
\author{Vladimir A. Zyuzin}
\affiliation{Department of Physics and Astronomy, Texas A$\&$M University, College Station, Texas 77843-4242, USA}
\affiliation{Nordita, KTH Royal Institute of Technology and Stockholm University,Roslagstullsbacken 23, SE-106 91 Stockholm, Sweden}
\begin{abstract}
We study Berry curvature driven and Zeeman magnetic field dependent electric current responses of two-dimensional electron system with spin-orbit coupling. 
New non-dissipative component of the electric current occurring in the applied in-plane magnetic field is described. 
This component is transverse to the electric field, odd in the magnetic field, and depends only on one particular direction of the magnetic field defined by the spin-orbit coupling. 
We show that the effect can be observed in a number of systems with $C_{3v}$ symmetry.
\end{abstract}
\maketitle

In the Hall effect \cite{Hall} the electric current acquires a component transverse to the applied electric field when the magnetic field is applied to the system. This is due to the Lorentz force conduction electrons experience and under which their trajectories get bent in the transverse direction. Thus the Hall effect happens in a configuration in which electric ${\bf E}$  and magnetic ${\bf B}$ fields and the current ${\bf j}$ are mutually orthogonal to each other \cite{Hall, Ziman}, and the current is proportional to the first power of the magnetic field, 
\begin{align}\label{Hall}
{\bf j}_{\mathrm{H}} \propto \left[ {\bf B} \times {\bf E} \right].
\end{align}

Another type of Hall effect, the anomalous Hall effect \cite{KarplusLuttinger,AHE_RMP}, occurs either in metals with magnetic order (ferromagnets or anti-ferromagnets) or in systems with applied Zeeman magnetic field. 
The effect does not originate from the Lorentz force but rather from the Berry curvature \cite{Berry} generated anomalous correction to the velocity, or due to special scattering processes which are also of geometric origin \cite{CulcerMacDonaldNiuPRB2003, Sinitsyn, BerryReview}. 
The underlying mechanism of the effect is the momentum-spin locking provided by the spin-orbit coupling \cite{Dresselhaus,Vas'koBychkovRashba, Dyakonov}, which as a result generates the Berry curvature.
In all known scenarios of anomalous Hall effect, just like in regular Hall effect, the three vectors: current, electric field, and magnetization ${\bf M} $ or Zeeman magnetic field are mutually orthogonal to each other,
\begin{align}\label{AHE}
{\bf j}_{\mathrm{AHE}} \propto \left[ {\bf M} \times {\bf E} \right].
\end{align}
For example, the conventional model \cite{CulcerMacDonaldNiuPRB2003} of the anomalous Hall effect consists of a two-dimensional electron system with Rashba spin-orbit coupling \cite{Vas'koBychkovRashba} and perpendicular to the plane magnetization or Zeeman magnetic field \cite{AHE_RMP}.

\begin{figure}[t] 
\centerline{
\begin{tabular}{cc}
\includegraphics[width=0.45\columnwidth,height=0.135\textheight]{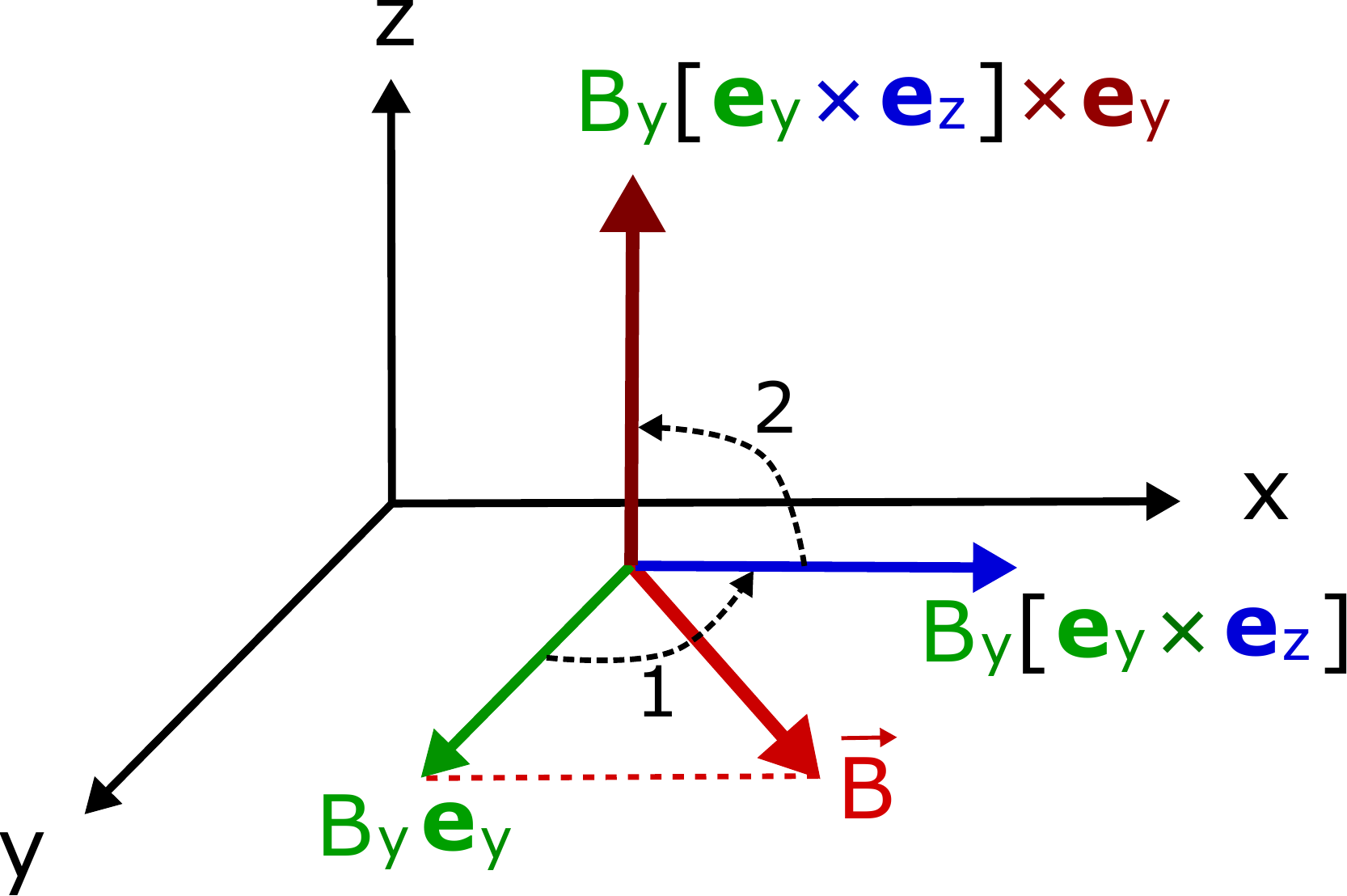}~~~&
\includegraphics[width=0.45\columnwidth,height=0.135\textheight]{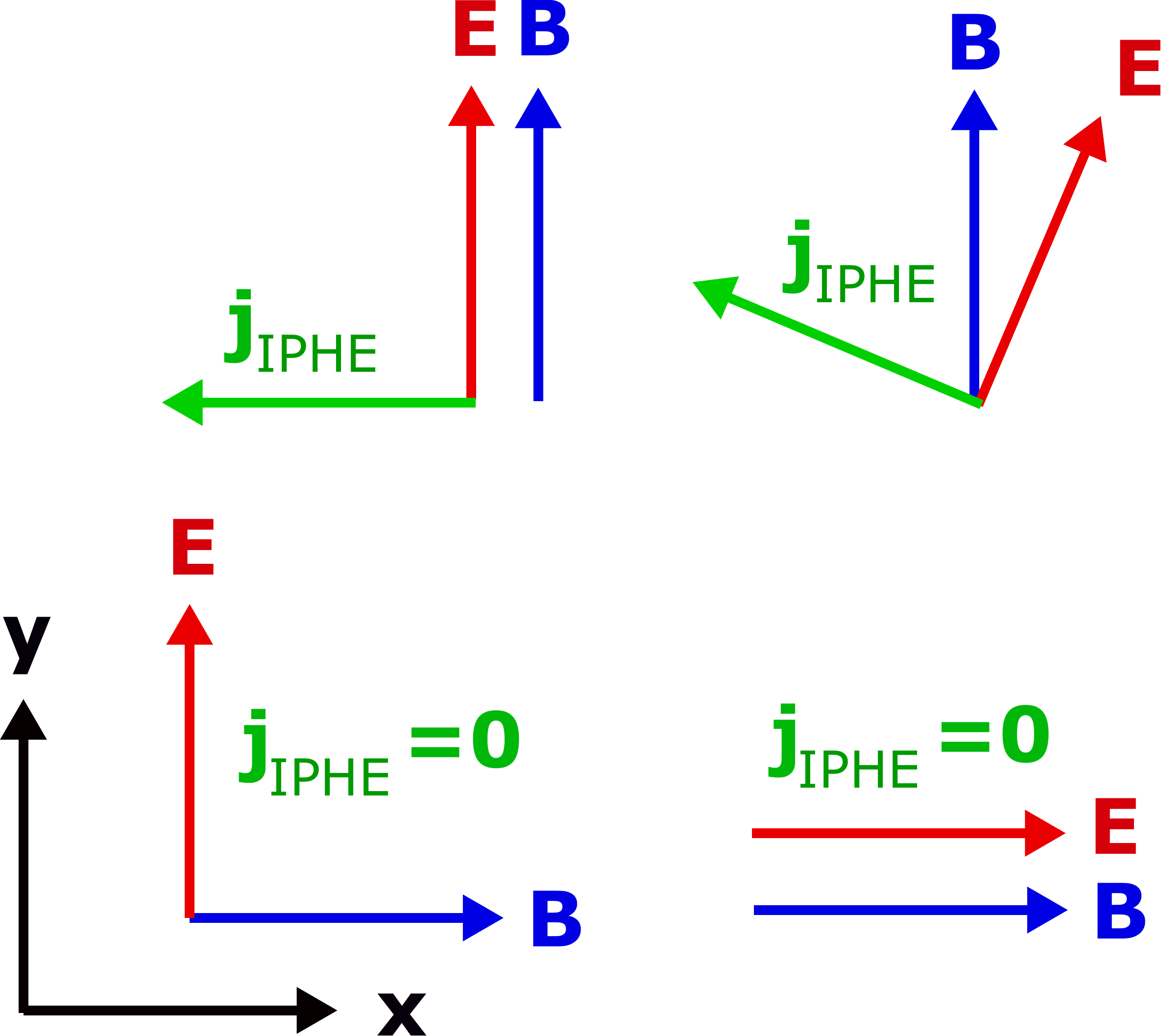} \\
\end{tabular}
}

\protect\caption{Left: Schematics of the lifting of the magnetic field. The first operation does $[{\bf B} \times {\bf e}_{\mathrm{z}}]$ to the $B_{y}$ component of the field, the second operation performs $\left[[{\bf B} \times {\bf e}_{\mathrm{z}}]\times {\bf e}_{\mathrm{y}}\right]$ making the overall vector normal to the plane. Only the $y-$ projection of the magnetic field is lifted, hence the choice of $B_{y}$, while the $x-$ projection will get filtered out in the process. 
Right: Schematics of the IPHE. All three vectors are in the $x-y$ plane. Only the $B_{y}$ component is needed to drive the in-plane Hall current. 
Dependence on angle between $B_{y}{\bf e}_{y}$ and the electric field ${\bf E}$ is of no importance.}

\label{fig1}  

\end{figure}

In this paper we demonstrate that the Berry curvature driven transverse, Hall-like, current can exist in two-dimensional electron system even when the Zeeman magnetic field and the electric field are parallel to each other. 
This is because the effective magnetic field effectively gets lifted from the $x-y$ plane of the system with the help of spin-orbit coupling, and the Hall response becomes proportional to the 
\begin{align}\label{ipheschematics}
{\bf j}_{\mathrm{IPHE}} \propto  \left[ \left[{\bf B}\times {\bf e}_{z}  \right] \times {\bf e}_{y}\right]\times {\bf E}
\end{align}
vector product, rather than to the conventional expression Eq. (\ref{Hall}).  
Here vectors ${\bf e}_{z}$ and ${\bf e}_{y}$ are due to the underlying spin-orbit coupling of the system.
As can be seen, only the $y-$ component of the magnetic field is responsible for the transverse current, and the angle between magnetic and electric fields is of no importance in this case. 
As a consequence, an unusual property of the effect is a configuration in which the electric field and magnetic field are set parallel to each other in $y-$ direction, but yet there is a transverse component of the current.
See Fig. (\ref{fig1}) for schematics.
Because of these reasons we propose to call this effect as the in-plane Hall effect (IPHE).

An obvious advantage of the IPHE, as compared to the mentioned conventional model of the anomalous Hall effect \cite{CulcerMacDonaldNiuPRB2003}, is the absence of possible Hall responses due to the Lorentz force. 
We propose two-dimensional $\sqrt{3}\times\sqrt{3}\mathrm{Au}/\mathrm{Ge}(111)$ \cite{HopfnerPRL2012}, $\mathrm{BiTeI}$ \cite{CrepaldiPRL2012, Fulop2018}, $\mathrm{BiAg}_{2}/\mathrm{Ag}/\mathrm{Si}(111)$ \cite{CredaldiPRB2012}, $\mathrm{Bi}(111)$ \cite{MiyamotoPRB2018}, and $\mathrm{Bi}_{2}\mathrm{Te}_{3}$ \cite{AlpichshevPRL, FuPRL2009} as possible candidates where IPHE can be observed. All of these materials have the $C_{3v}$ symmetry of the crystal structure. We show that in these materials IPHE is proportional to the third power of the magnetic field. Moreover, reduction of $C_{3v}$ symmetry to $C_{1v}$ with a stress makes the IPHE linear in the magnetic field. This is another unique experimental signature of the IPHE.

\textit{In-plane Hall effect}.-
To demonstrate the effect, we first study a model of $(110)$-grown quantum well, namely
a two-dimensional electron system with Rashba spin-orbit coupling and linear in momentum spin-orbit coupling allowed by the $C_{1v}$ symmetry. The Hamiltonian (for example, see \cite{CartoixaPRB2005}) of the system is,
\begin{align}\label{Hamiltonian}
\hat{H} = \frac{k^2}{2m}+\hat{H}_{\mathrm{s}} - \mu,
\end{align}
where $\mu$ is the chemical potential, and $H_{\mathrm{s}}$ is the spin part of the Hamiltonian given by
\begin{align}\label{SpinPart}
\hat{H}_{\mathrm{s}} = \sigma_{z} \Delta_{\bf k}+ v_{\mathrm{R}}(\sigma_{x} k_{y} - \sigma_{y} k_{x}) - \bf{h}\cdot{\bm \sigma},
\end{align}
where $\sigma_{x},\sigma_{y},\sigma_{z}$ are the Pauli matrices representing the electron's spin, ${\bf h} = \frac{1}{2}g\mu_{\mathrm{B}}{\bf B}$ is the Zeeman magnetic field chosen to lie in the plane, namely ${\bf h} = (h_{x},h_{y},0)$, $g$ is the $g-$factor, and $\mu_{\mathrm{B}}$ is the Bohr magneton, and $v_{\mathrm{R}}$ describes the Rashba spin-orbit coupling term \cite{Vas'koBychkovRashba}.
For $(110)$-grown quantum wells the Dresselhaus \cite{Dresselhaus} term is $\Delta_{\bf k} = v_{\mathrm{D}}k_{x}$ where $v_{\mathrm{D}}$ is a constant denoting the spin-orbit coupling strength.
The energy spectrum is 
\begin{align}
\epsilon_{{\bf k},\pm} = \frac{k^2}{2m}\pm\sqrt{ \Delta_{\bf k}^2 +  \vert \tilde{\gamma}_{\bf k} \vert^2  } - \mu,
\end{align}
where
$\vert \tilde{\gamma}_{\bf k} \vert^2 = (v_{\mathrm{R}} k_{y} - h_{x})^2+ (v_{\mathrm{R}}k_{x} + h_{y})^2$.
We will be using $v_{\mathrm{R}}\tilde{k}_{x} = v_{\mathrm{R}} k_{x} + h_{y}$ and $v_{\mathrm{R}}\tilde{k}_{y} = v_{\mathrm{R}} k_{y} - h_{x}$ notations in the following. 
Corresponding spinor wave functions are
\begin{align}
\Psi_{{\bf k},+} = \left[\begin{array}{c}
 \cos\left( \frac{\xi_{\bf k}}{2}\right) e^{i\chi_{\bf k}} \\
 - \sin\left( \frac{\xi_{\bf k}}{2}\right) \end{array} \right], ~
\Psi_{{\bf k},-} = \left[\begin{array}{c}
 \sin\left( \frac{\xi_{\bf k}}{2}\right) e^{i\chi_{\bf k}} \\
 \cos\left( \frac{\xi_{\bf k}}{2}\right) \end{array} \right],
\end{align}
where $\cos(\xi_{\bf k}) = \frac{\Delta_{\bf k}}{   \sqrt{ \Delta_{\bf k}^2 + \vert \tilde{\gamma}_{\bf k} \vert^2}}$, and $\chi_{\bf k} = \arctan\left(\frac{v_{\mathrm{R}}\tilde{k}_{y} }{v_{\mathrm{R}}\tilde{k}_{x}}\right)$ is the phase.

Intrinsic, non-dissipative, response of the system to static electric field ${\bf E}$ is given by an integral over the Berry curvature (can be derived within linear response formalism),
\begin{align}\label{current}
{\bf j}_{\mathrm{IPHE}} = e^2
\left[ \int\frac{d {\bf k}}{(2\pi)^2}\sum_{n = \pm} {\bm \Omega}^{(n)}_{\bf k} 
f(\epsilon_{{\bf k},n} ) \right]\times {\bf E},
\end{align}
where $f(\epsilon_{{\bf k},n} )$ is the Fermi-Dirac distribution function, and where we set $\hbar \equiv 1$ for the time being.
Note that there must also be dissipative responses which result in, for example, the Drude conductivity. 
We omit them in the following, as in our case they are not expected to result in transverse responses.
Berry curvature ${\bm \Omega}^{(n)}_{\bf k}$ for the two-dimensional system is in the $z-$ direction only, and its general expression is 
\begin{align}
&
\Omega^{(\pm)}_{z;\bf k}  =
2\mathrm{Im} \left(\partial_{k_{x}}\Psi^{\dag}_{{\bf k},\pm} \right) \left(\partial_{k_{y}}\Psi_{{\bf k},\pm} \right)
\\
&
=
\mp \frac{v_{\mathrm{R}}^2}{2\left( \Delta_{\bf k}^2 + \vert\tilde{\gamma}_{\bf k}\vert^2 \right)^{3/2}}
\left( \Delta_{\bf k} - \tilde{k}_{x}\partial_{x}\Delta_{\bf k} - \tilde{k}_{y}\partial_{y}\Delta_{\bf k}\right).
\nonumber
\end{align}
Calculations show that, for systems described by Hamiltonian Eq. (\ref{SpinPart}) only the $h_{y}$ contributes to the Berry curvature,
\begin{align}\label{ipheberry}
\Omega^{(\pm)}_{z;\bf k}
=
 \pm
\frac{v_{\mathrm{D}} v_{\mathrm{R}} h_{y} }
{2\left[  (v_{\mathrm{R}}\tilde{k}_{x})^2 + (v_{\mathrm{R}}\tilde{k}_{y})^2 + (v_{\mathrm{D}}k_{x})^2   \right]^{3/2}}.
\end{align}
Upon substituting the Berry curvature in to the definition of the current Eq. (\ref{current}), one gets for the in-plane Hall current,
\begin{align}\label{iphecurrent}
 {\bf j}_{\mathrm{IPHE}} 
=
\sigma_{\mathrm{IPHE}}
\left[ \left[{\bf e}_{\mathrm{B}}\times {\bf e}_{z}  \right] \times {\bf e}_{y}\right]\times {\bf E},
\end{align}
where ${\bf e}_{\mathrm{B}} = \frac{{\bf B}}{\vert B\vert}$ is the direction of the magnetic field.
In the expression, ${\bf e}_{y}$ is due to the Dresselhaus spin-orbit coupling, which intuitively can be thought of as a $k_{z}=0$ part of the $[{\bm \sigma} \times {\bf k}]_{y}$ product, while ${\bf e}_{z}$ is due to the Rashba spin-orbit coupling $[{\bm \sigma} \times {\bf k}]_{z}$ product.

The IPHE crucially depends only on one particular direction of the magnetic field defined by the vector product, the $y$ component in our case. If ${\bf h}$ is in $x-$ direction, then the overall vector product is zero, because of the filtering property of both ${\bf e}_{z}$ and ${\bf e}_{y}$ in the product, and there is no transverse current.
This is also because $h_{x}$ can be removed from the Berry curvature by shifting the $k_{y}$ momentum.
See Fig. (\ref{fig1}) for schematics, and for a unique configuration when the electric and magnetic fields are parallel to each other, but yet there is a transverse current.

In $(110)$-grown quantum wells \cite{CartoixaPRB2005}, the Rashba spin-orbit coupling is small and both Rashba split bands are occupied, which means a large chemical potential $\mu$ as compared to other parameters $h_{y}, v_{\mathrm{R}}k_{\mathrm{F}} , v_{\mathrm{D}}k_{\mathrm{F}}$, where $k_{\mathrm{F}}=\sqrt{2m\mu}$. We then analytically estimate the $\sigma_{\mathrm{IPHE}}$ for two cases. 
When $h_{y}<v_{\mathrm{R}}k_{\mathrm{F}}$, the IPHE conductivity is estimated 
\begin{align}
\sigma_{\mathrm{IPHE}}
\approx  
-
\frac{e^2}{4\pi}
\frac{ v_{\mathrm{D}} h_{y}}{\sqrt{v_{\mathrm{R}}^{2} + v_{\mathrm{D}}^{2} } \mu}  ,
\end{align}
when $h_{y}>v_{\mathrm{R}}k_{\mathrm{F}}$ and $h_{y}>v_{\mathrm{D}}k_{\mathrm{F}}$,
\begin{align}
\sigma_{\mathrm{IPHE}}
\approx  
-
\frac{e^2}{4\pi}
\frac{ (v_{\mathrm{D}}k_{\mathrm{F}}) (v_{\mathrm{R}}k_{\mathrm{F}})   }{h_{y} \mu}.
\end{align}
Therefore, the IPHE is very small when both Rashba split bands are occupied.

\textit{2D helical electron systems with $C_{3v}$}. -
Here we demonstrate that two-dimensional systems with $C_{3v}$ symmetry, which, as a consequence, have spin-orbit coupling of the hexagonal warping type, show cubic in magnetic field IPHE.
There are a number of materials with such a spin-orbit coupling, e.g.
 $(111)-$ grown zincblende quantum well \cite{CartoixaPRB2005}, 
$\sqrt{3}\times\sqrt{3}\mathrm{Au}/\mathrm{Ge}(111)$ \cite{HopfnerPRL2012}, $\mathrm{BITeI}$ \cite{CrepaldiPRL2012, Fulop2018}, $\mathrm{BiAg}_{2}/\mathrm{Ag}/\mathrm{Si}(111)$ \cite{CredaldiPRB2012}, $\mathrm{Bi}(111)$ \cite{MiyamotoPRB2018}, and $\mathrm{Bi}_{2}\mathrm{Te}_{3}$ \cite{AlpichshevPRL, FuPRL2009}.
Hexagonal warping is described by $\Delta_{\bf k} = \frac{\alpha}{2}(k_{+}^3 + k_{-}^3)$ (for example, see \cite{CartoixaPRB2005}) in the Hamiltonian Eq. (\ref{SpinPart}), where $k_{\pm} = k_{x} \pm i k_{y}$. 
Using the prescription introduced above, we derive the Berry curvature
\begin{align}
&
\Omega_{z;{\bf k}}^{(\pm)} 
= \mp \frac{1}{2} \frac{v_{\mathrm{R}}^2}{\left( \Delta_{\bf k}^2 + \vert\tilde{\gamma}_{\bf k}\vert^2 \right)^{3/2}}
\\
&
\times\left[ -2\alpha k_{x}^3 + 6\alpha k_{x}k_{y}^2 
- 3 \frac{\alpha h_{y}}{v_{\mathrm{R}}} (k_{x}^2 - k_{y}^2) 
- 6 \frac{\alpha h_{x}}{v_{\mathrm{R}}}  k_{x}k_{y}    \right].
\nonumber
\end{align}
Integration over the angles shows that IPHE vanishes in the linear order of Zeeman magnetic field.
However, the IPHE is proportional to a third power of the magnetic field. 
When the Fermi energy $\mu$ is the largest parameter in the system, we can analytically estimate IPHE to be
\begin{align}\label{iphehexagonal1}
\sigma_{\mathrm{IPHE}} \approx - \frac{e^2}{4\pi}\frac{\alpha k_{\mathrm{F}}^{3}}{(v_{\mathrm{R}}k_{\mathrm{F}})^3 \mu} h_{y}(h_{y}^2 - 3h_{x}^2),
\end{align}
for the case when $v_{\mathrm{R}}k_{\mathrm{F}} > \alpha k_{\mathrm{F}}^{3}$, and
\begin{align}\label{iphehexagonal2}
\sigma_{\mathrm{IPHE}} \approx - \frac{e^2}{4\pi}  \frac{(v_{\mathrm{R}} k_{\mathrm{F}})^5 }{(\alpha k_{\mathrm{F}}^3) h_{y}^3 \mu} ,
\end{align}
for $h_{y} \gg h_{x}$, $h_{y} \gg \alpha k_{\mathrm{F}}^3$, and $h_{y} \gg v_{\mathrm{R}} k_{\mathrm{F}}$.  
Just like in systems with $C_{1v}$ symmetry, IPHE given by Eqs. (\ref{iphehexagonal1}) and (\ref{iphehexagonal2}) has the same dependence on the electric and magnetic fields as Eq. (\ref{ipheschematics}), namely it requires a non-zero $h_{y}$. In addition it vanishes at a special value of $h_{y}=\pm \sqrt{3}h_{x}$. 

\begin{figure}[t] 
\centerline{
\begin{tabular}{cc}
\includegraphics[width=0.5\columnwidth,height=0.135\textheight]{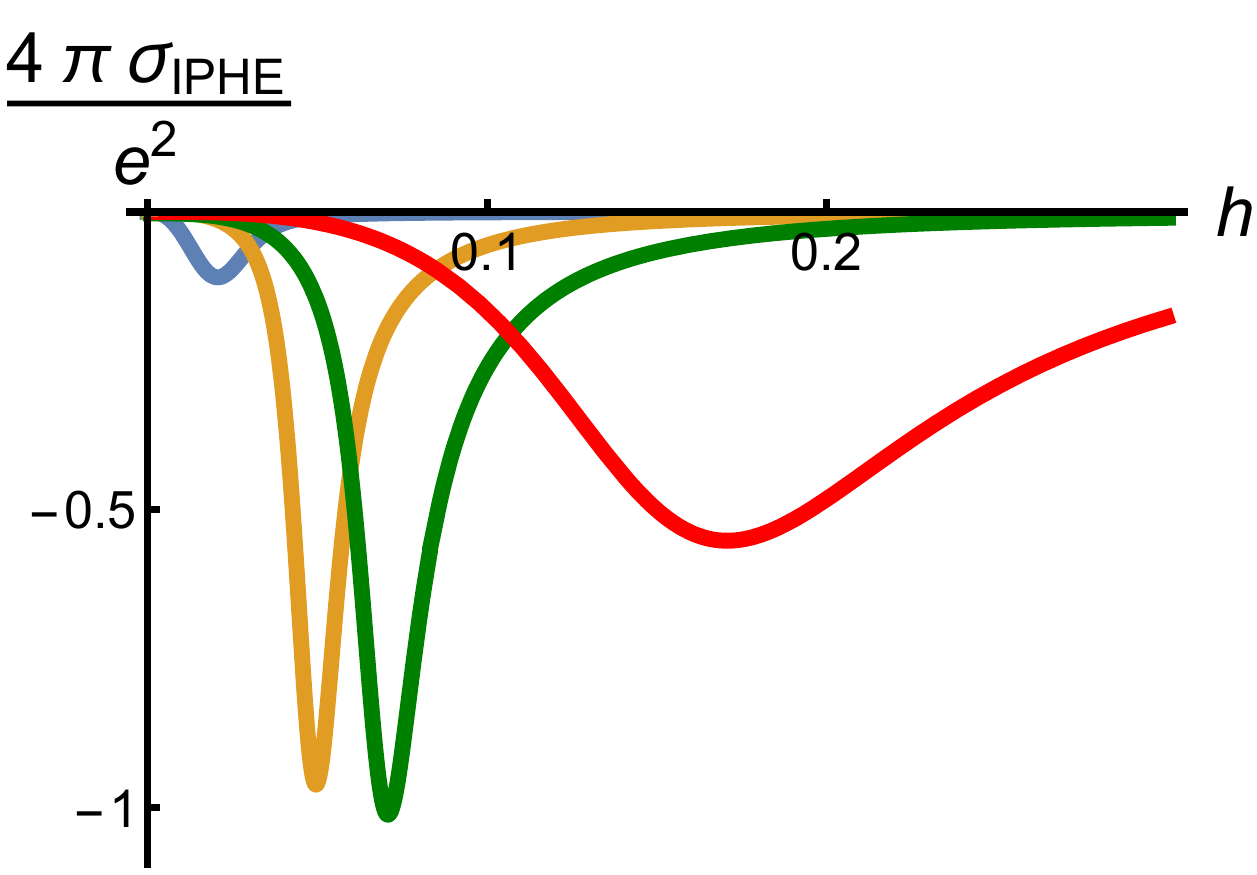}~&
\includegraphics[width=0.5\columnwidth,height=0.135\textheight]{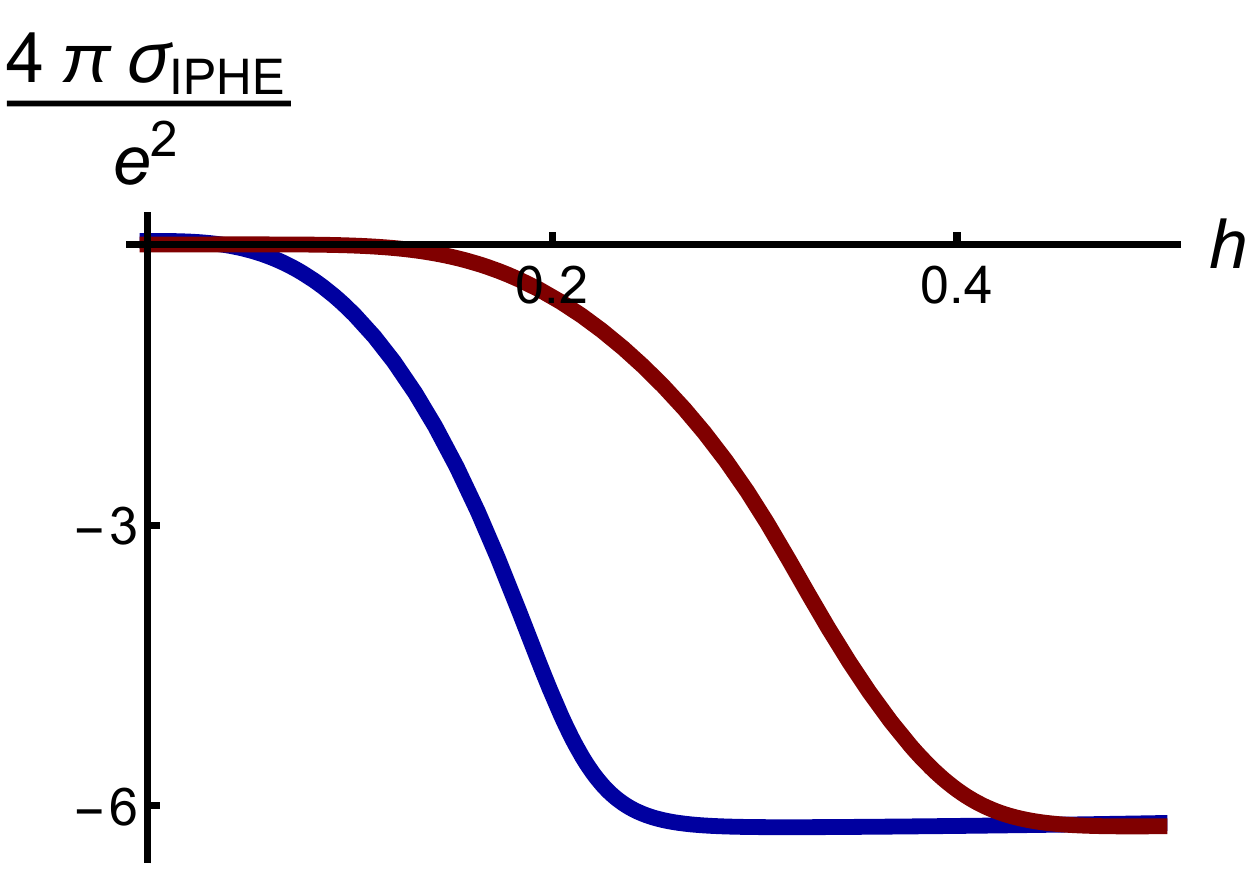} \\
\end{tabular}
}

\protect\caption{
Left: IPHE conductivity in $\mathrm{BiTeI}$ material with $C_{3v}$ symmetry with $\alpha = 150 \mathrm{eV}\AA^3$ and $v_{\mathrm{R}}= 2~ \mathrm{eV} \AA$ plot as a function of Zeeman magnetic field $h_{y}$ (in eV). 
Blue corresponds to $\mu = 0 $, yellow to $\mu=0.1 ~\mathrm{eV}$, green to $\mu=0.2 ~\mathrm{eV}$ and red $\mu =1~\mathrm{eV}$.
Right: IPHE conductivity in $\mathrm{Bi}_{2}\mathrm{Te}_{3}$ with $C_{3v}$ symmetry with $\alpha = 250\mathrm{eV}\AA^3$ and $v_{\mathrm{R}}= 2~ \mathrm{eV} \AA$ for blue $\mu=0.1 \mathrm{eV}$ and purple $\mu=0.2 \mathrm{eV}$. For the sake of numerical calculations, in both plots the temperature was taken to be $k_{\mathrm{B}}T=0.01~ \mathrm{eV}$, and $h=2\pi\hbar\equiv 1$. }

\label{fig2}  

\end{figure}

Despite smallness of the Eqs. (\ref{iphehexagonal1}) and (\ref{iphehexagonal2}), one can measure the IPHE in diluted semiconductors. It seems that from the list of the materials above, such are all but the first one. 
For example, in the left part of Fig. (\ref{fig2}) we plot the IPHE conductivity as a function of magnetic field at $h_{x}=0$ in $\mathrm{BiTeI}$ material with $m=0.003~\frac{1}{\mathrm{eV} \AA^2}$, $v_{\mathrm{R}} = 2 ~\mathrm{eV}\AA$, and $\alpha = 150~\mathrm{eV}\AA^3$ parameters for different values of Fermi energy. We see the cubic dependence of IPHE at small magnetic fields, and also conclude that the effect can be sizable.

In the right part of Fig. (\ref{fig2}) we plot IPHE conductivity in the topological insulator $\mathrm{Bi}_{2}\mathrm{Te}_{3}$ with $\frac{1}{m} = 0$ set in Eq. (\ref{Hamiltonian}), $v_{\mathrm{R}} = 2 ~\mathrm{eV}\AA$ and $\alpha = 250~\mathrm{eV}\AA^3$ for different values of chemical potential as a function of magnetic field at $h_{x}=0$. Again, the dependence of the effect on the magnetic field is cubic at small fields, but it saturates at a value of $\sigma_{\mathrm{IPHE}} = -\frac{e^2}{2}$ as the magnetic field is increased. The growth corresponds to the case when both bands are occupied, while the plateau to the case when the chemical potential is in the gap. Quantization of the AHE in this material was proposed in Ref. \cite{LiuPRL2013, ZhangPRB2019}.

\begin{figure}[t] 
\centerline{
\begin{tabular}{cc}
\includegraphics[width=0.5\columnwidth,height=0.135\textheight]{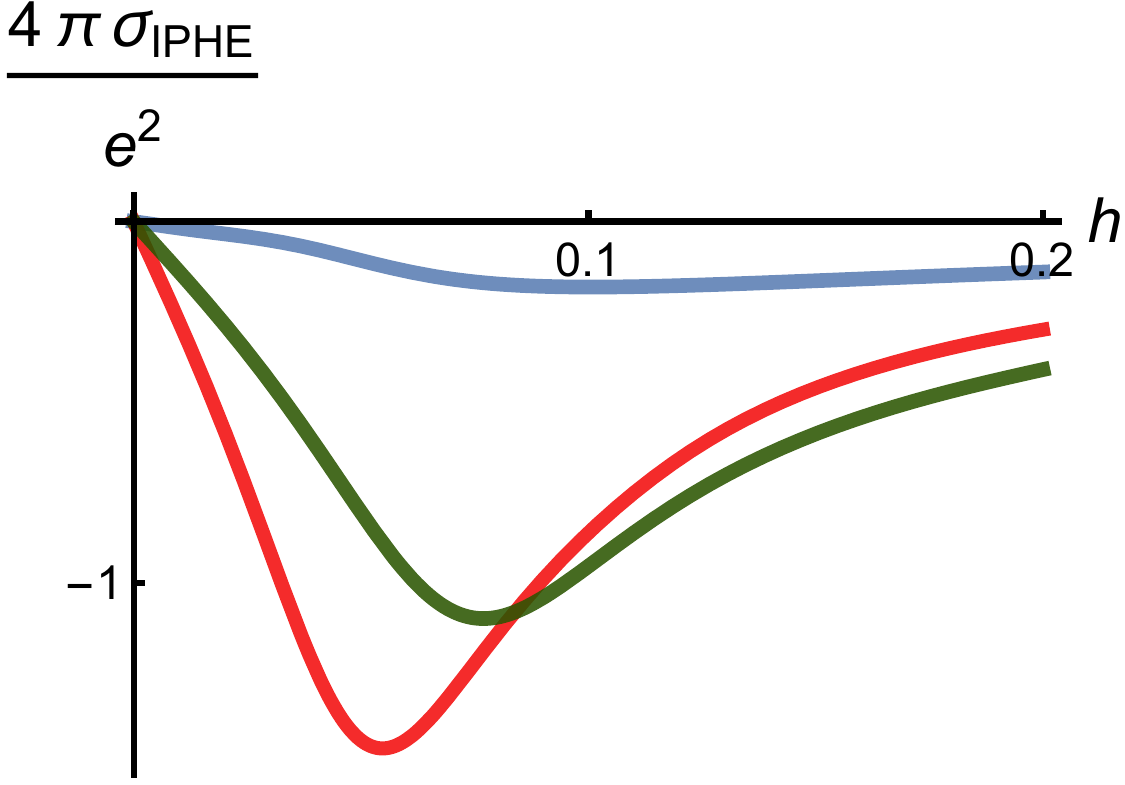}~&
\includegraphics[width=0.5\columnwidth,height=0.135\textheight]{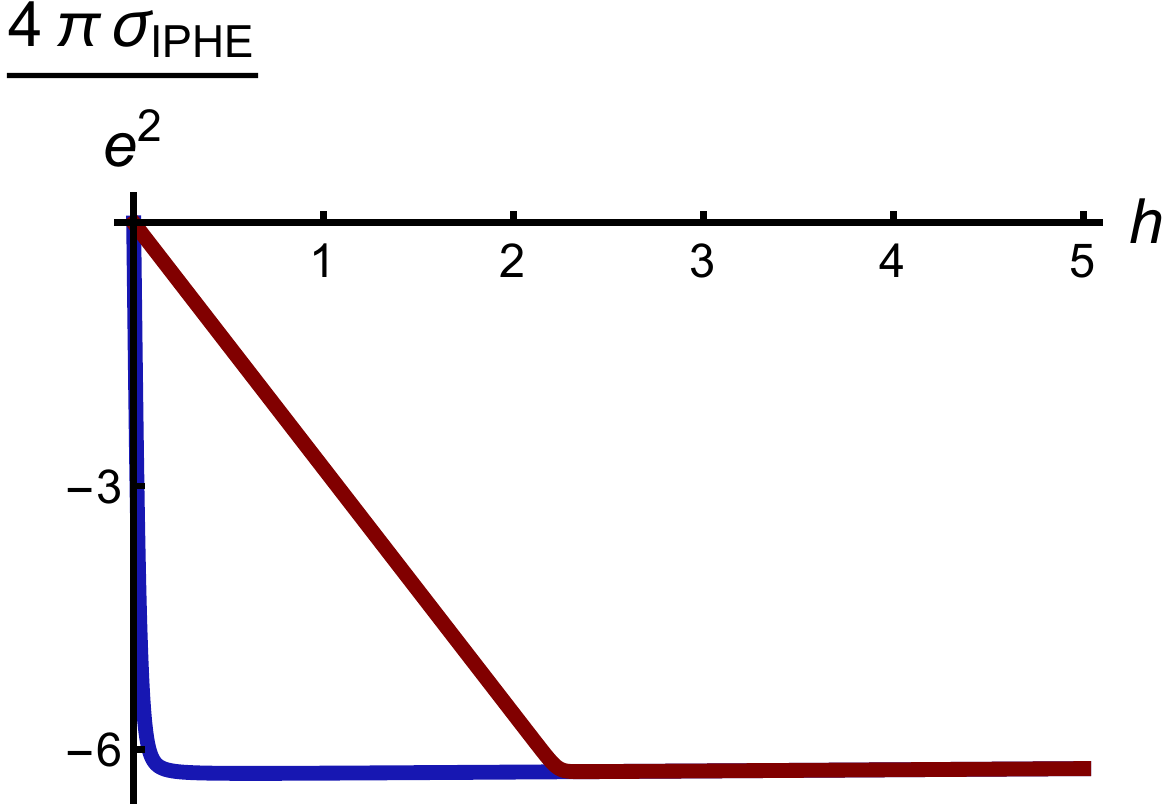} \\
\end{tabular}
}

\protect\caption{
Left:
IPHE conductivity in $\mathrm{BiTeI}$ material with $C_{1v}$ symmetry (with $\alpha = 0$) plot as a function of Zeeman magnetic field $h_{y}$ (in eV). 
When $\mathrm{BiTeI}$ is stressed the $v_{\mathrm{D}}$ spin-orbit coupling was taken to be $v_{\mathrm{D}} = 2 ~\mathrm{eV}\AA$. Blue corresponds to $\mu = -0.05 ~\mathrm{eV}$, red to $\mu=0.1 ~\mathrm{eV}$ and green $\mu =0.2~\mathrm{eV}$.
Right: IPHE conductivity in a model of topological insulator $\mathrm{Bi}_{2}\mathrm{Te}_{3}$ with $C_{1v}$ symmetry (with $\alpha = 0$) plot as a function of magnetic field $h_{y}$ (in eV). 
Assumed parameters are $v_{\mathrm{R}}= 2~ \mathrm{eV} \AA$ and $v_{\mathrm{D}}= 1~ \mathrm{eV} \AA$, blue $\mu = 0$ and purple $\mu_{1}=1~ \mathrm{eV}$. 
For the sake of numerical calculations, in both plots the temperature was taken to be $k_{\mathrm{B}}T=0.01~ \mathrm{eV}$, and $h=2\pi\hbar\equiv 1$. 
}

\label{fig3}  

\end{figure}

\textit{Reducing $C_{3v}$ symmetry to $C_{1v}$}.-
As suggested in \cite{ZhangLiuPRB2019}, one can reduce the $C_{3v}$ symmetry to $C_{1v}$ by applying a stress on the system. 
Thus, the materials with $C_{3v}$ symmetry discussed above will be described by a Hamiltonian Eq. (\ref{SpinPart}) with $\Delta_{\bf k}=v_{\mathrm{D}}k_{x} + \frac{\alpha}{2}(k_{+}^3 + k_{-}^3)$. Then, according to calculations leading to Eqs. (\ref{ipheberry}) and (\ref{iphecurrent}), there will be a component in the IPHE linear in the magnetic field. To explicitly show that, we discuss the case when $\alpha =0$.

In the left part of Fig. (\ref{fig3}) we plot the IPHE conductivity as a function of magnetic field in the ${\mathrm{BiTeI}}$ material with the parameters given above, for different values of Fermi energy. We observe that the effect is now linear in the magnetic field at small fields. 
In the right part of Fig. (\ref{fig3}) the same is plotted for the $\mathrm{Bi}_{2}\mathrm{Te}_{3}$ topological insulator. When both bands are occupied [purple plot in right part of Fig. (\ref{fig3})], the conductivity is linear in the magnetic field. When the magnetic field is such that the chemical potential is in the gap, the conductivity becomes quantized, $\sigma_{\mathrm{IPHE}} = -\frac{e^2}{2}$.

When both $\alpha \neq 0$ and $v_{\mathrm{D}} \neq 0$, we expect behaviors of Figs. (\ref{fig2}) and (\ref{fig3}) to mix with each other. For example, cubic low-field dependence in Fig. (\ref{fig2}) will be replaced with linear, while the $\propto \frac{1}{h_{y}}$ decay in the left plot of Fig. (\ref{fig3}) will be replaced with $\propto\frac{1}{h_{y}^3}$.

\textit{Discussion}.-
Let us compare the IPHE, Eq. \ref{ipheschematics}, with the known cases.
Recall, that in the ordinary Hall effect the current, magnetic field and electric field are mutually orthogonal, ${\bf j}_{\mathrm{H}} \propto [{\bf E}\times {\bf B} ]$, while in the IPHE only the current and electric field are orthogonal and the magnetic field is in the plane with them.
To draw an analogy between the IPHE and the ordinary Hall effect, one can say that in the former case with the aid of spin-orbit coupling the magnetic field is lifted from the plane to become normal to the plane. 
Note that, while ${\bf h}$ is in $y-$ direction, vector product $\left[ \left[{\bf h}\times {\bf e}_{z}\right] \times {\bf e}_{y}\right]$ is in $z-$ direction, thus making the derived current contribution resemble the Hall effect. 
Moreover, in contrast with the in-plane transverse magnetoconductivity, i.e., when $\delta {\bf j}$, ${\bf E}$, and ${\bf B}$ are all in one plane and by transverse we mean $\delta {\bf j}\perp {\bf E}$, for example corresponding component of $\delta{\bf j} \propto ({\bf E}\cdot{\bf B}){\bf B}$ (for a review see \cite{Ziman, SeitzPR1950,GoldbergDavisPR1954,KyJETP1966} and Ref. \cite{CommentA}, in context of Weyl semimetals see Refs. \cite{zrte5_exp1,Yip, ZyuzinWSM} and Ref. \cite{CommentB}), where the current is proportional to the second power of the magnetic field, the derived effect explained in this paper effect is linear in the magnetic field.

We note that there is no scattering life time $\tau$ in the expression for the derived current, as we have derived only the intrinsic part to the Hall current. 
However, there must be disorder contributions to the effect. 
We have to point out that in Ref. \cite{Mal'shukovChaoWillander} some disorder effects have already been studied for (110) grown quantum well. However, we are skeptical about their results because of their obtained divergent $v_{\mathrm{R}} \rightarrow 0$ limit, hexagonal warping contributing linear magnetic field dependence without vanishing at $h_{y} = \pm \sqrt{3}h_{x}$, and $\frac{1}{k_{\mathrm{F}}\ell}$ smallness ($\ell$ is the electron's mean free path) of the effect. Despite that, Ref. \cite{Mal'shukovChaoWillander} captures the Eq. (\ref{ipheschematics}) dependence.
In established theory \cite{Sinitsyn, AHE_RMP} the disorder is known to contribute to the anomalous Hall effect with the side-jump and skew-scattering processes. These extra contributions are expected to be described \cite{Sinitsyn} by the same field dependence as in the Eq. (\ref{ipheschematics}).  

In the known studied Rashba spin-orbit coupling and out-of-plane Zeeman magnetic field model \cite{CulcerMacDonaldNiuPRB2003}, the anomalous Hall effect vanishes due to the mutual cancellations of the intrinsic and disorder contributions only when the two Rashba split bands are both occupied \cite{NunnerPRB2007}. 
Thus the above discussed $\mathrm{BiTeI}$ monolayer and $\mathrm{Bi}_{2}\mathrm{Te}_{3}$ (and possibly the others mentioned above) are good candidates to avoid such a possibility in the IPHE. We believe that, before attempting tedious \cite{Sinitsyn, AHE_RMP} calculations of disorder effects, a future experiment should first shed light on the fate of the IPHE in these two materials, keeping present paper's predictions as a starting point.

We note that photogalvanic effects were studied \cite{IvchenkoPikus,IvchenkoGanichev, Bel'kov} in the systems with spin-orbit coupling.
There, in particular, magneto-gyrotropic effects were observed, where the applied Zeeman magnetic field, together with spin-orbit coupling, result in unusual non-linear in electric field but linear in magnetic field current responses (for example, $\delta j_{x} \propto \chi h_{x}(E_{x}^2-E_{y}^2) + 2\chi h_{y} E_{x}E_{y}$, where $\chi$ is a constant odd in spin-orbit coupling strength \cite{IvchenkoPikus}). Hopefully, in light of this, the IPHE can be observed within existing experimental techniques.

The IPHE discussed here essentially has the same nature as the anomalous Hall effect. 
In Ref. \cite{ZZ2015} it was shown that, if the system has the anomalous Hall effect, there will be a possibility to excite chiral electromagnetic waves\cite{ZhukovRaikhPRB2000} in it. 
In Ref. \cite{ZZ2015} such waves were proposed to occur in magnetized Weyl semimetals. 
The chiral electromagnetic waves will propagate at the boundary of the system or at the domain wall between two opposite Zeeman magnetic field orientations in only one particular direction - they are uni-directional. 
They are carried by electronic edge modes which are there in any system with Hall effect, and, technically, restore gauge invariance of the bulk Maxwell equations. We therefore expect chiral waves to propagate in the systems studied in this paper.
The chiral waves should be contrasted with the chiral magnetoplasmon waves \cite{Seshadri1962, Fetter1985} due to orbital effects of electrons.
The in-plane magnetic field will make the chiral waves easier to observe, as the orbital effects of electrons in this case are absent.

Finally, we would like to contrast the discussed IPHE with the model of anomalous Hall effect in two-dimensional electron systems with Rashba spin-orbit coupling and out-of-plane Zeeman magnetic field \cite{CulcerMacDonaldNiuPRB2003}. 
The latter case might be hard to reach experimentally as one will have a challenge of separating the anomalous Hall effect from the regular, due to the Lorentz force, Hall effect. 
In the IPHE scenario one can apply strong in-plane magnetic field which will only act on an electron's spin and barely any orbital motion will be present. 
The effect should also occur in three-dimensional ferromagnets, where orbital effects are absent. 
Hopefully descriptions made in this paper will guide further research activities to identify such ferromagnets and/or materials with peculiar spin-orbit coupling.

\textit{Conclusions}.-
To conclude in this paper we have described a new non-dissipative transverse component of the current given by Eq. (\ref{ipheschematics}). 
Its origin, just like in the anomalous Hall effect, is due to the Berry curvature. 
This current has a peculiar dependence on the magnetic field shown in Fig. \ref{fig1}, namely it is driven by only one component of the field, $B_{y}$ in our case. As long as the magnetic field is pointing in $y-$ direction, there will be a transverse current regardless of the mutual orientation of electric and magnetic fields. We proposed to call the effect as the in-plane Hall effect.
In all discussed examples of material candidates, the IPHE can be sizable and might be experimentally observed (see Figs. \ref{fig2} and \ref{fig3}).

The author is thankful to M.M. Glazov, A.A. Kovalev, and P.O. Sukhachov for discussions, to A. G. Mal'shukov for pointing out his work \cite{Mal'shukovChaoWillander} to us after the release of this paper on arxiv, and to Pirinem School of Theoretical Physics, where parts of this work were completed, for hospitality.  This work is supported by the VILLUM FONDEN via the Centre of Excellence for Dirac Materials (Grant No. 11744), the European Research Council under the European Unions Seventh Framework Program Synergy HERO, and the Knut and Alice Wallenberg Foundation KAW 2018.0104.

\end{document}